\documentclass[conference]{IEEEtran}
\IEEEoverridecommandlockouts
\usepackage{cite}
\usepackage{amsmath,amssymb,amsfonts}
\usepackage{algorithm}
\usepackage{algorithmic}
\usepackage{graphicx}
\usepackage{subfigure}
\usepackage{textcomp}
\usepackage{xcolor}
\usepackage[fontsize=10.05pt]{fontsize}
\def\BibTeX{{\rm B\kern-.05em{\sc i\kern-.025em b}\kern-.08em
    T\kern-.1667em\lower.7ex\hbox{E}\kern-.125emX}}
\begin{document}

\title{A Dynamic Subarray Structure in Reconfigurable Intelligent Surfaces for TeraHertz Communication Systems}
% % \thanks{Identify applicable funding agency here. If none, delete this.}
% }

\author{\IEEEauthorblockN{ Yicong Liu}
\IEEEauthorblockA{\textit{School of Engineering and IT } \\
\textit{The University of Sydney}\\
Sydney, Australia \\
yicong.liu@sydney.edu.au}
\and
\IEEEauthorblockN{ Weijie Li}
\IEEEauthorblockA{\textit{School of Engineering and IT } \\
\textit{The University of Sydney}\\
Sydney, Australia \\
weli0332@uni.sydney.edu.au}
\and
\IEEEauthorblockN{ Zihuai Lin}
\IEEEauthorblockA{\textit{School of Engineering and IT } \\
\textit{The University of Sydney}\\
Sydney, Australia \\
zihuai.lin@sydney.edu.au}
}

\maketitle

\begin{abstract}
Reconfigurable Intelligent Surface (RIS) has become a popular technology to improve the capability of a THz multiuser Multi-input multi-output (MIMO) communication system. THz wave characteristics, on the other hand, restrict THz beam coverage on RIS when using a uniform planar array (UPA) antenna. In this study, we propose a dynamic RIS subarray structure to improve the performance of a THz MIMO communication system. In more details,  an RIS is divided into several RIS subarrays according to the number of users. Each RIS subarray is paired with a user and only reflects beams to the corresponding user. Based on the structure of RIS, we first propose a weighted minimum mean square error - RIS local search (WMMSE-LS) scheme, which requires that each RIS element has limited phase shifts. To improve the joint beamforming performance, we further develop an adaptive Block Coordinate Descent(BCD)-aided algorithm, an iterative optimization method. Numerical results demonstrate the effectiveness of the dynamic RIS subarray structure and the adaptive BCD-aided joint beamforming scheme and also show the merit of our proposed system.
\end{abstract}


\section{Introduction}
TeraHertz (THz) communication is becoming increasingly popular in wireless networks as a means of improving data transmission rates. THz communications provide a substantially larger capacity than millimeter wave (mmWave) and microwave communications. However, the THz wave has some shortages in practice, such as a much higher path loss, and the obstacles can easily block the line-of-sight (LOS) path. %\cite{chen2019survey,petrov2016terahertz,lin2015indoor,xu2021THzUAV,xu2022THzRIS}. 
\cite{xu2021THzUAV,xu2022THzRIS}. 
As a result, the Reconfigurable Intelligent Surface (RIS) is increasingly being used in THz communications to improve the non-line-of-sight (NLOS) route \cite{liu2021reconfigurable,Ding2016Nlos,UAV_2,UAVdownlink}. Since the RIS is a passive device with low energy consumption and without self-interference, it is regarded as a better technology than the backscatter and Multi-input multi-output (MIMO) relay \cite{chu2022IoT_RIS,Hu2021CodedRIS,hu2021Backscatter,Chen2021MIMO,hu2019Ambc,xing2018ambc,MIMO_capacity,network_capacity,fountaincodes2014}. In a RIS-aided THz communication system, the Base Station (BS) and users usually have a uniform planar array (UPA) antenna to improve the data transmission rates. More antennas and RIS elements also can provide a higher rate. However, when the number of UPA antennas and RIS elements increases, the system sum-rate will increase, but RIS utilization will reduce because the THz beam is more narrow and hard to cover all RIS elements.  

Many existing works have tried to improve the beam coverage from BS to RIS. In \cite{zhai2020thzprism}, an 'THzPrism' scheme was proposed to extend the angle of THz beams. The results showed that the bandwidth would lower down but still be sufficient during the increase of angular coverage. On another side, the research of \cite{chen2019sum} proposed a cross-entropy method to improve the joint beamforming performance in THz, reducing the algorithm complexity. The results showed that when the RIS elements increase with fixed BS size, the weighted sum-rate (WSR) growth rate becomes meager which means that the number of RIS elements can not grow infinitely. The authors of \cite{ge2021ris} attempted to determine the optimal number of UPA antennas and RIS components as a compromise between hardware cost, algorithm complexity, and system performance. Aside from that, beamforming is an important part of beam coverage. The beam coverage would be improved if the optimal combined beamforming between BS and RIS was near to perfect. When the number of RIS elements grows, the algorithm complexity typically grows as well \cite{siddiqi2021low}. In \cite{guo2020weighted}, the authors proposed an optimized Block coordinate descent (BCD) scheme to maximize the system sum-rate and reduce computing cost, which can counteract the impact of RIS elements increase.

In this study, we propose a dynamic RIS subarray structure in a THz MIMO system, with the number of subarrays changing in response to the number of users. Each RIS subarray only serves one corresponding user, different from the general RIS structure. To keep pace with this special structure in RIS, we first construct an environment for the RIS-aided THz system using the 3GPP standard \cite{3gpp2018study}. Since the THz communications rapidly attenuates as it goes through the obstacles, the LOS path is assumed to be blocked while there is only NLOS path in direct channel \cite{bai2013coverage}. In addition, both BS and users are equipped with UPA antennas, which is a tendency in THz because the active beamforming performance of UPA antennas is better than Uniform Linear Array(ULA) in same conditions \cite{chatterjee2017evaluation}. In joint beamforming processing among BS, RIS, and users, we first propose a weighted minimum mean square error - RIS local search (WMMSE-LS) scheme where LS is for RIS and WMMSE is applied at BS. The number of available phase shifts of RIS is finite in the LS method, limiting RIS passive beamforming accuracy. To improve the joint beamforming performance, we propose an adapted BCD-aided algorithm, where the beamforming matrices of BS and RIS are updated iteratively with no limitation in RIS phase shifts. Simulation results demonstrate that both the dynamic RIS subarray structure and adapted BCD-aided joint beamforming can improve the performance of the RIS-aided THz communication system.

Notations: {$\mathbf{I_{\textit{$N$}}}$} is a $N \times N $ identity matrix. {$\mathbf{(\cdot)^*}$}, {$\mathbf{(\cdot)^T}$}, and {$\mathbf{(\cdot)^H}$} denote conjugate, transpose, and conjugate transpose of matrix. {$Re(\cdot)$} is the real part of the complex number. {$Im(\cdot)$} is the imaginary part of the complex number.

\section{System models}
In our MIMO communication system model, we mainly consider the downlink from the BS to users via the RIS or the RIS subarrays where the BS generates beams and align beams to the users to maximize the sum-rate of the network. The joint beamforming in this system includes active beamforming on BS and users, and passive on RIS/RIS subarrays.
\begin{figure}[htbp]
\centering
\subfigure[RIS without subarray]
{\label{Fig.without subarray} \includegraphics[width=\linewidth]{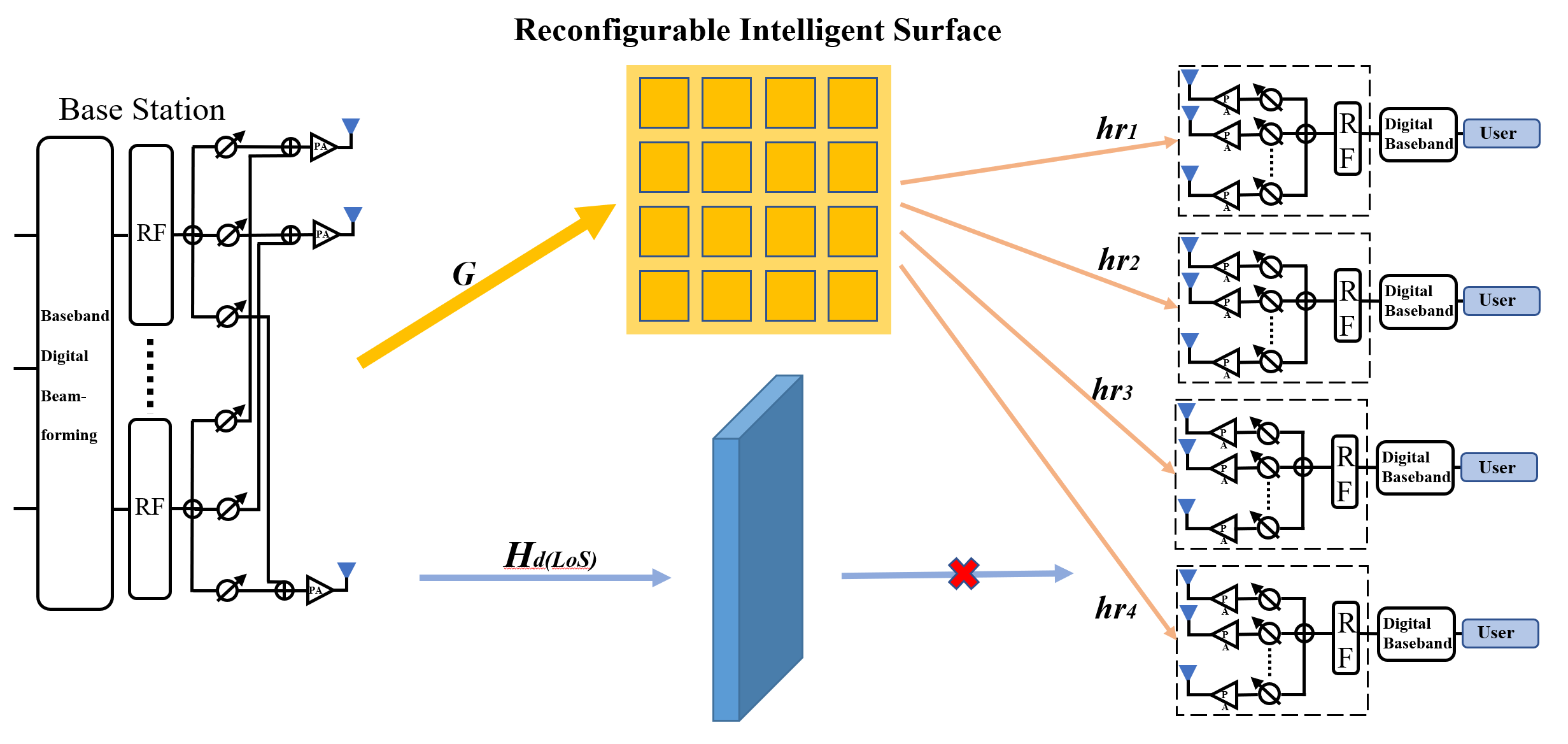}}
\subfigure[Dynamic RIS subarray structure]
{\label{Fig.dynamic subarray} \includegraphics[width=\linewidth]{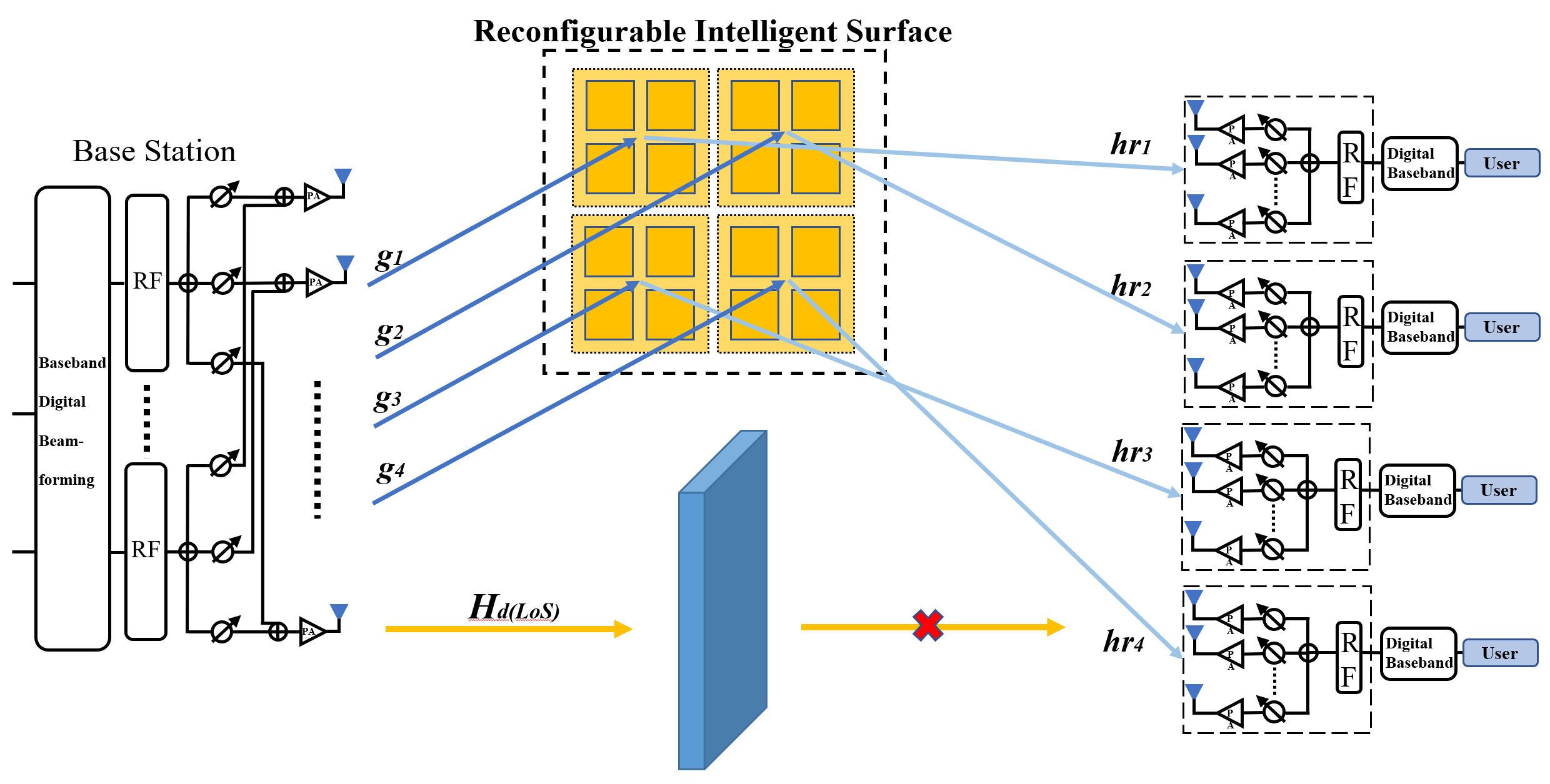}}
\caption{System model with or without RIS subarray (Users $J=4$)}
\label{fig:system model}
\end{figure}

As shown in Fig. \ref{fig:system model}, there are two different RIS structures in our system model where the number of users $J$ is assumed to be 4. Fig. \ref{Fig.without subarray} indicates a system where the RIS is a general type without subarray, and Fig. \ref{Fig.dynamic subarray} demonstrates a dynamic RIS subarray structure. In this system, the BS is equipped with  {$N_{t} \times M_{t}$} UPA antenna elements, and the RIS is an {$N \times N$} array servicing $J$ users where each user has {$N_{r} \times M_{r}$} UPA antennas. There are $P$ subarrays in the dynamic RIS subarray structure, and we assume that $P=J$ in the proposed system. Since each user is paired with one RIS subarray, the received vector at the $j$-th user can be expressed as:
% \begin{small}
\begin{subequations}
\begin{equation}
    \mathbf{y_{j}}= \mathbf{v_{j}^H(h_{d,j}^H+h_{r,p,j}^H\Theta_{p}g_{p})}\sum_{i=1}^{J}\mathbf{w}_{i}\mathbf{s}_{i}+\mathbf{n}
\end{equation}
\begin{equation}
    \mathbf{\Theta_{p}=diag}[\theta_{\frac{(p-1)N^{2}}{P}+1},\theta_{\frac{(p-1)N^{2}}{P}+2},\cdots,\theta_{\frac{pN^{2}}{P}}]
\end{equation}
\end{subequations}
% \end{small}\\
where $\mathbf{w}_{i}$ is the transmitting beamforming vector of BS to the $i$-th user, $\mathbf{s}_i$ is the transmission signal vector,  $\mathbf{\Theta_{p}}$ is the phase shift matrix of the $p$-th RIS, $\theta_{n}$ is the phase shift of the $n$-th RIS element, $\theta_{n}=e^{j\psi}$, and $\mathbf{v_{j}}$ is the receiving vector of the $j$-th user. Besides, $\mathbf{h_{d,j}^H}$ is the direct channel between BS and user without LOS path, {$\mathbf{g_{p}} \in \mathbb{C}^{\frac{N^{2}}{P} \times N_{t}M_{t}}$} is the channel between BS and $p$-th RIS subarray, and {$\mathbf{h_{r,p,j}} \in \mathbb{C}^{\frac{N^{2}}{P} \times N_{r}M_{r}}$}  is between the $p$-th RIS subarray and the $j$-th user. The $\mathbf{n}$ is the additive white Gaussian noise (AWGN). The reflecting channel follows the Rician channel model according to \cite{han2019large}, which can be expressed as:
% \begin{small}
\begin{equation}
{\mathbf{g_{p}}}={P_{Lg}}(\sqrt{\frac{k_{1}}{k_{1}+1}}\Bar{\mathbf{a_{1,p}}}+\sqrt{\frac{1}{k_{1}+1}}{\mathbf{\Tilde{g_{p}}}})
\end{equation}

\begin{equation}
\Bar{ \mathbf{a_{1,p}}}=\mathbf{a_{sub-RIS}}(\varphi_{AoA,RIS},\phi_{AoA,RIS})\mathbf{a_{BS}^{H}}(\varphi_{AoD,BS},\phi_{AoD,BS})
\end{equation}

\begin{equation}
 \mathbf{h_{r,p,j}^{H}}={P_{Lr,j}}(\sqrt{\frac{k_{2,p}}{k_{2}+1}}\Bar{\mathbf{a_{2}}}+\sqrt{\frac{1}{k_{2}+1}}{\mathbf{\Tilde{h_{r,p,j}}}})
\end{equation}

\begin{equation}
\Bar{\mathbf{a_{2,p}}}=\mathbf{a_{j}}(\varphi_{AoA,j},\phi_{AoA,j})\mathbf{a_{sub-RIS}^{H}}(\varphi_{AoD,j},\phi_{AoD,j})
\end{equation}
% \end{small}\\
where $P_{L}$ is the pathloss of channels and $k$ is the Rician factor, ${\mathbf{\Tilde{g_{p}}}}$ and ${\mathbf{\Tilde{h_{r,p,j}}}}$ are NLOS path components of $\mathbf{g_{p}}$ and $ \mathbf{h_{r,p,j}^{H}}$, respectively.     

The directional vector $\mathbf{a}$ can be expressed as:
\begin{equation}
\begin{split}
\mathbf{a_{UPA}(\varphi,\phi)}= &[1,\cdots,e^{j2\pi\frac{d}{\lambda}(m\sin{\varphi}\sin{\phi}+n\cos{\phi})},\\
&\cdots,e^{j2\pi\frac{d}{\lambda}((W-1)\sin{\varphi}\sin{\phi}+(H-1)\cos{\phi})}]^{T}\\
\end{split}
\end{equation}\\
where $\varphi$ and $\phi$ are the angles in the respective horizontal and vertical directions, $W$ and $H$ are the numbers of elements in the  horizontal and vertical directions, respectively. 

\section{Joint beamforming scheme}
The sum-rate and Signal to Interference and Noise Ratio (SINR) of the system are given below.
% \begin{small}
\begin{subequations}
    \begin{equation}
    R=\sum_{j=1}^{J} \omega_{j} \log_{2}(1+\gamma_{j})
    \end{equation}
    \begin{equation}
    \gamma_{j}=\frac{{\lvert \mathbf{v_{j}^{H} (h_{d,j}+h_{r,p,j}^{H} \Theta_{p} g_{p}) w_{j}}\lvert}^{2} }{{\sum_{i=1,i\neq j}^{J}\lvert \mathbf{v_{j}^{H}(h_{d,j}+h_{r,p,j}^{H} \Theta_{p} g_{p})} \mathbf{w}_{i}\lvert}^{2}+{\sigma}^{2}}
    \end{equation}
\end{subequations}
% \end{small}\\
where $\omega$ is the weight of one user, which is determined by the pathloss between the BS and this user. The whole channel $\mathbf{h_{j}}$ between the BS and the $j$-th user is $\mathbf{h_{j}=h_{d,j}+h_{r,p,j}^{H} \Theta_{p} g_{p}}$.

\subsection{WMMSE-LS method}
The key to maximize the sum-rate is to find appropriate beamforming vectors of the BS, RIS, and users. Due to the limited computing capability and hardware complexity, the number of available vectors is finite in one system generally. We first assume that the accuracy of each RIS element's phase shift is $r$-bit. Each RIS element has $2^r$ available phase shifts, and the whole RIS has $2^{(r \times N^2)}$ available vectors. In addition, we also presume the accuracy of users' UPA antenna is $q_1$-bit and $q_2$-bit in the horizontal and vertical direction, respectively. 

% In this work, we first propose a WMMSE-LS algorithm to determine rapidly the beamforming vectors of RIS and users. The phase shift of RIS elements is searched in sequence independently. 

In order to quickly determine the beamforming vectors of RIS and users, we first propose a WMMSE-LS algorithm in this work. The phase shift of RIS elements is individually and sequentially searched.
For instance, while searching the phase shift of the $n$-th RIS element, the previous $n-1$ elements have been searched and fixed. The phase shift of the remaining undetermined RIS elements are randomly generated from the phase shift set. 
When the number of RIS components rises, the overhead of exhaustive searching becomes very substantial, necessitating the search of all $2^{(r \times N^2)}$ possible phase shift matrices. As a result, compared to the exhaustive technique, the search space for the RIS phase shift adjustment only has to explore  $2^r \times N^2$ phase shift matrices.
%
% Because the overhead of exhaustive searching will be extremely high, which requires search all $2^{(r \times N^2)}$ possible phase shift matrix, when the number of RIS elements increases. Based on that, in the RIS phase shift adjusting, it only requires searching $2^r \times N^2$ phase shift matrices, being significantly smaller than the exhaustive approach.
%11
The set of possible phase shifts for RIS is {$\mathcal{F}_{LS}$}, and the sets of angles for each user are {$\mathcal{F}_{q_{1}} $= $ \{ 1, \frac{\pi}{q_{1}},\cdots,\frac{n\pi}{q_{1}},\cdots,\pi \} $} in the horizontal plane and {$\mathcal{F}_{q_{2}} $= $ \{-\frac{\pi}{2},  (\frac{\pi}{q_{2}}-\frac{\pi}{2}),\cdots,(\frac{n\pi}{q_{2}}-\frac{\pi}{2}),\cdots,\frac{\pi}{2} \} $} in the vertical plane.
   
Determining each user's beamforming vector is the first step in this algorithm. The searching step starts from the first user and the first beamforming angle in the $\mathcal{F}_{LS}$. While searching is ongoing, the beamforming angles of the remaining users are determined randomly. The vector of the $j$-th user is $\mathbf{v_{j}}= \frac{1}{\sqrt{N_{r}M_{r}}}\mathbf{a_{j}}(\varphi_{q_{1},j},\phi_{q_{2},j})$. After the users' beamforming vectors are determined, to simplify the algorithm description, we define an equivalent channel between BS and users, which follows that $\mathbf{\check{h_{j}}=v_{j}^{H}h_{j}}$. Then RIS will search the available phase shifts, and the BS will align beams by using the WMMSE method \cite{shi2011iteratively}. In Algorithm 1, the WMMSE-LS joint beamforming algorithm is illustrated.

\begin{algorithm}[h]
\caption{WMMSE-LS method}
\label{alg:LS}
    \begin{algorithmic}[1]
    \REQUIRE Quantified phase shifts, {$r$}-bit
    \REQUIRE Receiving beamforming vector $\mathbf{V}$ from all users
    \ENSURE  RIS phase shift matrix {$\mathbf{\Theta}$}, BS's beamforming {$\mathbf{W}$}
    \STATE   Initialize {$\mathcal{F}_{LS} = \{ e^{j0},e^{j\frac{2\pi}{2^{r}}},\cdots,e^{j(2^{r}-1)\frac{2\pi}{2^{r}}} \}$}
    \FOR{\textit{$n=1:N^2$}}
    \STATE  Randomly generate {$\{ \theta_{n+1},\cdots,\theta_{N^2}\} \in \mathcal{F}_{LS}$}
    \FOR{\textit{$n_{r}=1:2^{r}$}}
    \STATE  {$\theta_{n}=\mathcal{F}_{LS}(n_{r})$}
    \STATE  {$\mathbf{\Theta_{n_{r}}=diag[\theta_{1},\theta_{2},\cdots,\theta_{\textit{$N^2$}}]}$}
    \STATE  Generate equivalent channel {$\mathbf{\check{H}}$}
    \STATE  Calculate {$\mathbf{W_{n_{r}}}$},{$\mathbf{w_{j}}$} for all users by WMMSE
    \STATE  Calculate sum-rate {$R_{n_{r}}$},{$R_{j}$}
    \ENDFOR
    \STATE  Find out and save {$R_{n,max}$} ,{$\theta_{n,max}$}, and {$\mathbf{W}_{n,max}$}
    \STATE  Save {$\theta_{n}=\theta_{n,max}$}
    \ENDFOR
    \STATE   Find out and save {$R_{max}$} ,{$\theta_{max}$}, and {$\mathbf{W_{max}}$}
    \end{algorithmic}
\end{algorithm}

\subsection{Adaptive BCD-aided joint beamforming}
Considering the RIS devices manufacturing, the accuracy of RIS elements is not very high generally. Despite that, the joint beamforming algorithm can be further improved without considering these limitations. In this paper, we propose an enhanced BCD aided joint beamforming method, in which the phase shift configuration of RIS elements has no accuracy limitations. 

The proposed algorithm is described below. Based on the closed-form scheme in \cite{guo2020weighted}, the sum-rate and its maximization can be formulated as:
% \begin{small}
    \begin{equation}
    \begin{split}
        &\textit{$f_{opt}$} (\alpha,\beta,\mathbf{W,\Theta}) =
        \sum_{j=1}^{J} \textit{$f_{j,opt}$} (\alpha_{j},\beta_{j},\mathbf{w_{j},\Theta_{j}})\\
        &=\sum_{j=1}^{J}(\omega_{j}(\log_2(1+\alpha_{j})-\alpha_{j})
        +2\sqrt{\omega_{j}(1+\alpha_{j})}Re(\beta_{j}^* \mathbf{\check{h_{j}}w_{j}})\\
        &-{\lvert \beta_{j} \lvert}^{2}(\sum_{i=1}^{J} {\lvert \mathbf{\check{h_{j}}}\mathbf{w}_{i} \lvert}^{2}+\sigma^{2}))\\
    \end{split}
    \label{equ:W-update}
    \end{equation}
% \end{small}
The optimization variables {$\alpha_{j}$} and {$\beta_{j}$} can be expressed by \cite{xu2013block}:
% \begin{small}
\begin{subequations}
    \begin{equation}
        \alpha_{j}=\frac{{\Bar{\eta_{j}}}^2+{\Bar{\eta_{j}}}\sqrt{{\Bar{\eta_{j}}}^2+4}}{2}
    \end{equation}
   \begin{equation}
    \beta_{j}=\frac{\sqrt{\omega_{j}(1+\Bar{\alpha_{j}})}(\mathbf{\Bar{\check{h_{j}}}\Bar{w_{j}}})}{\sum_{i=1}^{J}{\lvert \mathbf{\Bar{\check{h_{j}}}}\mathbf{\Bar{w}}_{i} \lvert}^{2}+\sigma^2}
    \end{equation}
\end{subequations}
% \end{small}\\
where ${\Bar{\eta_{j}}}=\frac{1}{\sqrt{\omega_{j}}}Re(\Bar{\beta_{j}}^* \mathbf{\Bar{\check{h_{j}}}\Bar{w_{j}}})$.  {$\Bar{\alpha},\Bar{\beta},\mathbf{\Bar{w_{j}}}$}, and {$\mathbf{\Bar{\Theta}}$} are temporal optimized results from last iteration. {$\mathbf{\Bar{\check{h_{j}}}}$} is the equivalent channel matrix generated from the previous iteration according to  {$\mathbf{\Bar{\Theta}}$} in this iteration.

When updating the RIS phase shift matrix $ \mathbf{\Theta=diag[\Theta_{1},\Theta_{2},\cdots,\Theta_{p},\cdots,\Theta_{P}]}$, where $\mathbf{\Theta_{p}}$ is the $p$-th RIS subarray which follows (1b), the phase shift of each RIS element can be generated as {$\theta_{n}=e^{j\psi}$}. $\mathbf{\Theta_{p}}$ is updated separately and the rest of subarray phase shifts keep fixed. When the RIS subarrays are paired with users, the dimensions of both phase shift matrices and channel matrices decrease, which greatly decrease the complexity. For instance, when updating the whole RIS in the general RIS model, the dimension of $\mathbf{\Theta}$ in each iteration is $N^2 \times N^2$, the BS-RIS channel $\mathbf{G}$ is $N_{t}M_{t} \times N^{2}$, and the RIS-user channel $\mathbf{h_{r,j}}$ is $N^{2} \times N_{r}M_{r}$. In the RIS subarray model, each iteration only optimizes the phase shift matrix $\mathbf{\Theta_{p}}$ of size $\frac{N^{2}}{P} \times \frac{N^{2}}{P}$  and the size of separate channels, $\mathbf{g_{j}}$ and $\mathbf{h_{r,j}}$, are $N_{t}M_{t} \times \frac{N^{2}}{P}$ and $\frac{N^{2}}{P} \times N_{r}M_{r}$. The reflecting channel is transfered as $\mathbf{h_{RIS,j}=diag(h_{r,j})G}$. $\psi$ updating follows as \cite{bertsekas1997nonlinear}:

\begin{subequations}
\begin{equation}
    \psi =\rm{argmin} \textit{$f_{\psi}$}(\psi)
    \triangleq \mathbf{(\textit{$e^{j\psi}$})^{H}A}\mathbf{(\textit{$e^{j\psi}$})}-2Re(\mathbf{(\textit{$e^{j\psi}$})^{H}B})
\end{equation}
    \begin{equation}
        \mathbf{A}=\sum_{j=1}^{J}{\lvert \beta_{j} \lvert}^2 \sum_{i=1}^{J}(\mathbf{h_{RIS,j}} \mathbf{\Bar{w}}_{i})(\mathbf{h_{RIS,j} }\mathbf{\Bar{w}}_{i})^{H}
    \end{equation}
    \begin{small}
    \begin{equation}
    \hspace{-4mm}
    \mathbf{B}=\sum_{j=1}^{J}(\sqrt{\omega_{j}(1+\Bar{\alpha_{j}})}\Bar{\beta_{j}}^{*}(\mathbf{h_{RIS,j}\Bar{w_{j}}})-{\lvert \beta_{j} \lvert}^2\sum_{i=1}^{J}(\mathbf{h_{d,j}}\mathbf{\bar{w}}_{i})^*(\mathbf{h_{RIS,j}}\mathbf{\Bar{w}}_{i}))
    \end{equation}
    \end{small}
\end{subequations}

After $\mathbf{\Theta}$ is updated and fixed, then update $\mathbf{W}$ on the BS. The prox-linear update rules \cite{xu2013block} of $\mathbf{W}$ can be expressed as:
% \begin{small}
\begin{subequations}
\begin{equation}
    \mathbf{W}=\rm{argmax} \sum_{j=1}^{J} (Re(\mathbf{Gra_{j}}^{H}(\mathbf{w_{j}-\hat{w_{j}}}))+\frac{\textit{$L$}}{2} {\Vert \mathbf{w_{j}-\hat{w_{j}}}\Vert}^{2})
\end{equation}
    \begin{equation}
        \mathbf{Gra_{j}} =-2\sqrt{\omega_{j}(1+\Bar{\alpha_{j}})}\Bar{\beta_{j}}\mathbf{\Bar{\check{h_{j}}}}+2\sum_{i=1}^{J}{\lvert \beta_{i} \lvert}^2 \mathbf{\Bar{\check{h}}}_{i}\mathbf{\Bar{\check{h}}}_{i}^{\mathbf{H}} \mathbf{\hat w_{j}}
    \end{equation}
\end{subequations}
% \end{small}\\
where  {$\mathbf{Gra}$} is the gradient matrix. $\mathbf{\hat{w_{j}}}$ follows $\mathbf{\hat{w_{j}}= \Bar{w_{j}}+ \textit{$\tau$} (\Bar{w_{j}}-\Ddot{w_{j}})}$. {$\mathbf{\Ddot{w_{j}}}$} is the value from last iteration, and {$\mathbf{\hat{w_{j}}}$} is the extrapolated point. {$\textit{L}$} follows $ L=2{\Vert\sum_{i=1}^{J}{\lvert \beta_{i}\lvert}^2\mathbf{\Bar{\check{h}}}_{i}\mathbf{\Bar{\check{h}}}_{i}^{H}\Vert}$. The Algorithm 2 briefly summarizes the procedure of this scheme.

\begin{algorithm}[h]
\caption{Adapted BCD-aided optimization algorithm}
    \label{alg:adapted BCD}
    \begin{algorithmic}[1]
    \REQUIRE Beamforming vector {$\mathbf{v_j}$}, Initialized {$\mathbf{\Theta_{p}^{(0)}}$} 
    \ENSURE Weighted sum-rate $R$
        \STATE  Randomly initialized {$\mathbf{\Theta_{p}^{(0)}}$}
        \STATE  Generate the equivalent channel {$\mathbf{\check{h_{p,j}}}$} for each user
        \STATE  Initialize {$\mathbf{W^{(0)}}$} and {$\mathbf{w_{j}}^{(0)}$} by WMMSE
        \STATE  Initialize {$\alpha^{(0)}$}, {$\alpha_{j}^{(0)}$}, {$\beta^{(0)}$}, {$\beta_{j}^{(0)}$}, and i=0 
        % \STATE  set i=0 
        \FOR{j=1:J}
        \REPEAT
        \STATE  Update {$\psi^{(i)}$}, {$\psi \in \mathbf{\Theta}$}
        \STATE  Update {$\beta_{j}^{(i)}$}
        \STATE  Update {$\mathbf{W^{(i)}}$} and {$\mathbf{w_{j}^{(i)}}$}
        \STATE  Update {$\alpha_{j}^{(i)}$}, {$\beta_{j}^{(i)}$}
        \STATE  i=i+1 
        \UNTIL  {The value of rate {$R_{j}$} follows \textit{$f_{j,opt}$} }
        \ENDFOR
    \end{algorithmic}
\end{algorithm}

\section{Numerical Results}
\subsection{Simulation scenario}
\begin{figure}[htbp]
\centering
{\includegraphics[width=\linewidth]{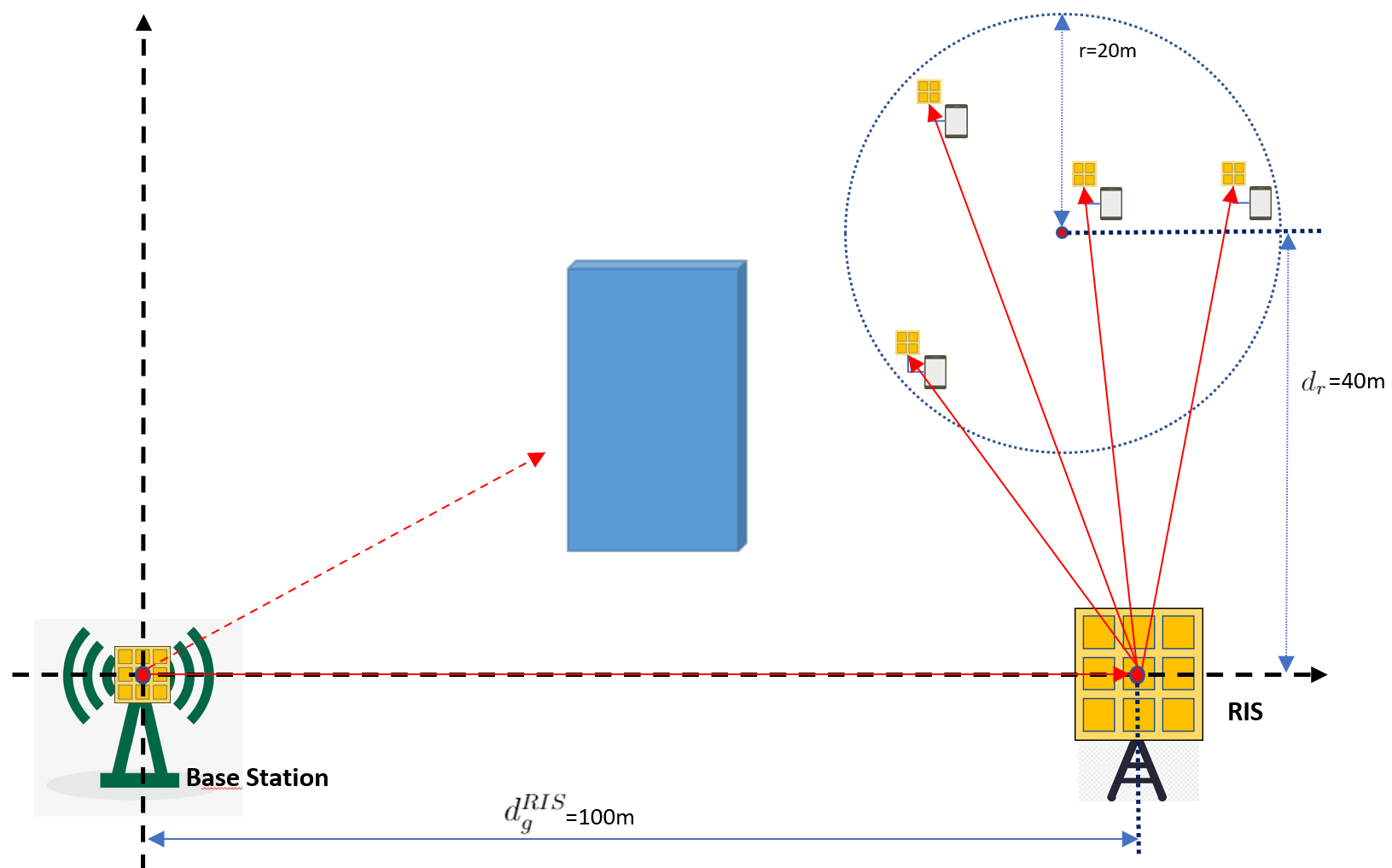}}
\caption{The simulation scenario}
\label{fig:simulation scenario}
\end{figure}

As shown in Fig. \ref{fig:simulation scenario}, since the LOS channel between the BS and users are blocked, a RIS is deployed to assist data communications. The distance between the BS and RIS is fixed, which is set as {$d_{g}^{RIS}=100m$}. {$d_{d,j}^{user}$} is used to denote the distance between the BS and the {$j$}-th user while {$d_{r,j}^{user}$} is the distance between RIS and the j-th user. The locations of users are randomly generated from a circle range, where the distance between the circle center and the RIS is set as {$d_{r}=40m$}, and the radius of the gathering zone is 20m. This communication model can be numerically presented in a cartesian coordinates system, where the BS is located at (0,0), RIS is located at (100, 0) and the center of the gathering zone is located at (100,40). The NLOS path of the direct channel follows Rayleigh fading model while the RIS reflecting cascade channel follows Rician fading, where the Rician factors in (2) and (4) are set as {$k_{1}=k_{2}=k_{2,p}=10$}. The related parameters are shown in Table. \ref{tab:parameters}.

\begin{table}[htbp]
\centering
\caption{Simulation Parameters}
\scalebox{1.0}{
\begin{tabular}{|c|c|}
\hline
Parameters & Values  \\
\hline
BS Location & (0m,0m) \\
\hline
RIS Location & (100m,0m)\\
\hline
Carrier Frequency {$f_{c}$} & 100 GHz \\
\hline
Transmission BandWidth & 10 Ghz\\
\hline
Noise power spectral density & -220 dBm/Hz\\
\hline
Path-loss for {$\mathbf{G/g_{j},h_{r,j}}$} (dB) & 32.4 + 21log10({$d$}) + 20log10({$f_{c}$}) \\
\hline
Path-loss for {$\mathbf{H_{d}/h_{d,j}}$} (dB) & 22.4 +35.3log10({$d$})+21.3log10({$f_{c}$}) \\
\hline
\end{tabular}}
\label{tab:parameters}
\end{table}

\subsection{Model parameters and results analysis}

\begin{figure}[htbp]
    \centering
    \includegraphics[width=\linewidth]{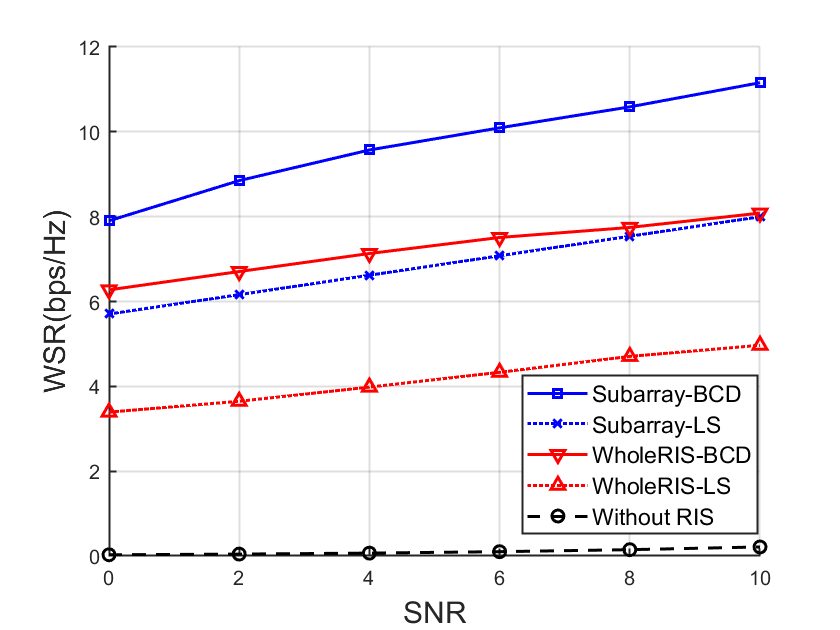}
    \caption{WSR versus SNR}
    \label{fig:WSR-SNR}
\end{figure}

As shown in Fig. \ref{fig:WSR-SNR}, when model parameters are set as {$N_{t}=M_{t}=N_{r}=M_{r}=2$}, {$N=10$}, {$J=4$}, {$q_{1}=2,q_{2}=1$}, {$r=1$}, the sum-rate of the adapted BCD-aided joint beamforming is higher than that of the WMMSE-RIS local search method at the same SNR. Meanwhile, the dynamic RIS subarray structure with both adapted BCD-aided and WMMSE-RIS local search can provide a higher sum-rate than the entire RIS for joint beamforming system, which indicates that the adapted BCD algorithm has a better performance compared with the dynamic RIS subarray structure.

\begin{figure}[H]  
\centering  
\subfigure[Rate of each user versus iteration times {$I$}(RIS Subarray)]
{   \label{fig:subarray-SNR=5}  \includegraphics[width=0.23\textwidth]{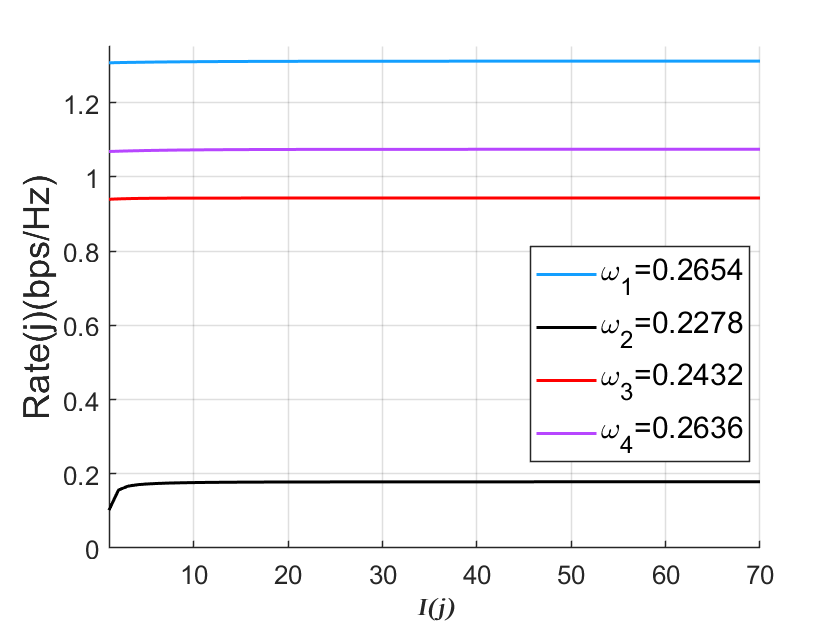}}  
\subfigure[WSR versus iteration times (RIS Subarray)]
{   \label{fig:subarray-sum}  \includegraphics[width=0.23\textwidth]{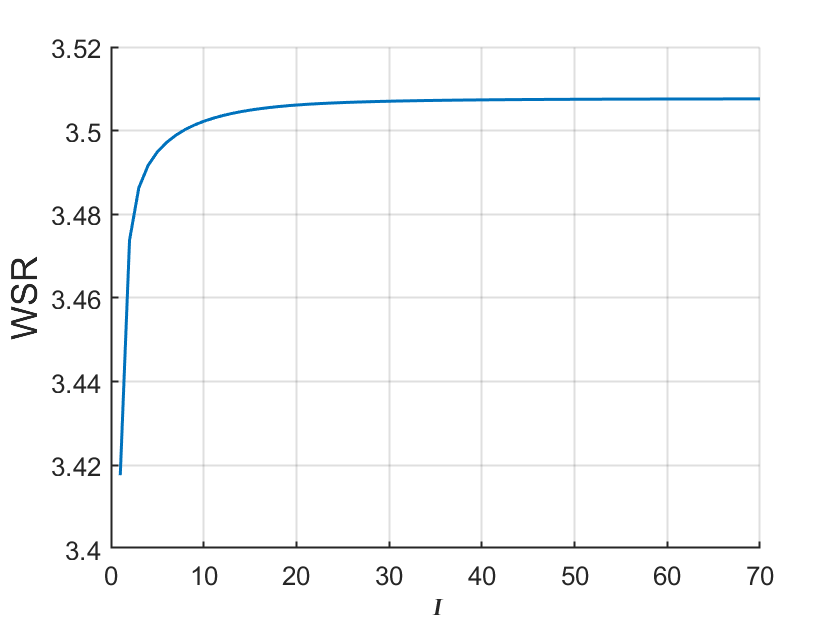}}  
\subfigure[WSR versus iteration times (Whole RIS)]
{   \label{fig:subarray-sum}  \includegraphics[width=0.25\textwidth]{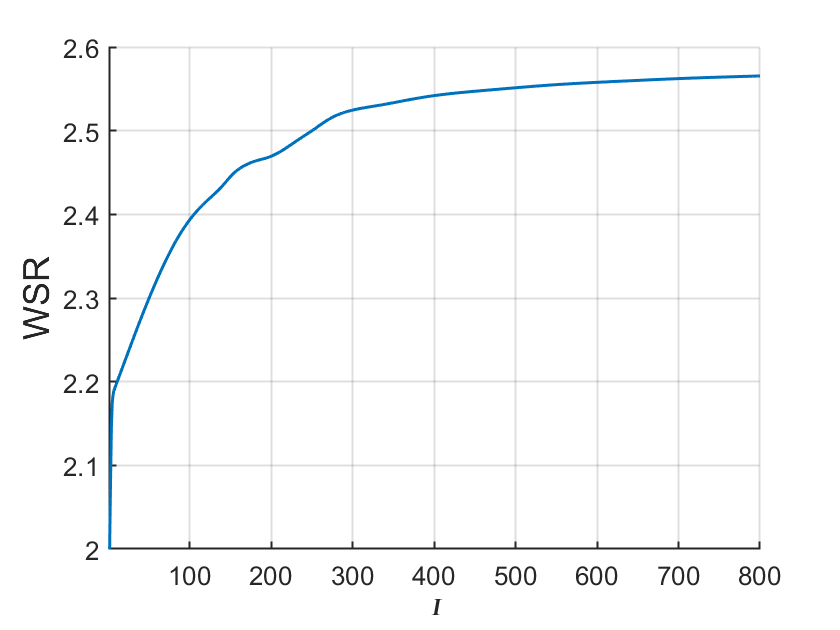}} 
\caption{Rate versus iteration times} 
\label{fig:WSR-I}  
\end{figure}

Fig. \ref{fig:WSR-I} describes the relationship between the rate and the number of iterations in the BCD optimization. 
As can be observed, when the transmitting SNR is 0dB, the rate increases as the number of iterations increases.  Although the improvement for each user is different, the entire system still converges quickly with the RIS subarray structure.  As the sum-rate improvement reaches convergence after the 50th iteration, the joint beamforming can be regarded as complete.

% It can be seen that with the increase of iterations, the rate rises up when the transmitting SNR is 0dB. Although the improvement for each user is different, the entire system still converges quickly with the RIS subarray structure. The joint beamforming can be considered as complete after the 50th iteration as the sum-rate improvement achieves convergence.

\begin{figure}[htbp]
    \centering
    \includegraphics[width=\linewidth]{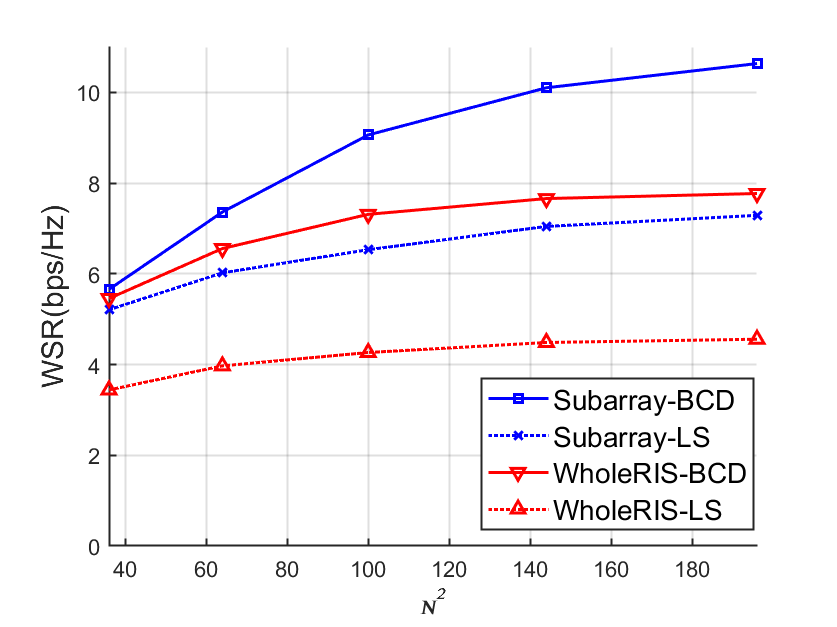}
    \caption{WSR versus RIS elements {$N^2$} (SNR=5)}
    \label{fig:WSR-Nl}
\end{figure}

Fig. \ref{fig:WSR-Nl} shows the system performance against the number of RIS elements where the channel parameters remain the same. It should be noted that the number of RIS elements {$N^2$} should be divisible by {$J$}. The results reveal that the proposed RIS subarray scheme outperforms the whole RIS scheme. 
% and as the number of RIS elements increases, the WSR slowly grows. One possible explanation is that although more RIS elements can provide greater improvement, the utilization of the marginal elements may be lower due to the limitation of beam coverage especially when the RIS is significantly larger than the BS. 

\begin{figure}[htbp]
    \centering
    \includegraphics[width=\linewidth]{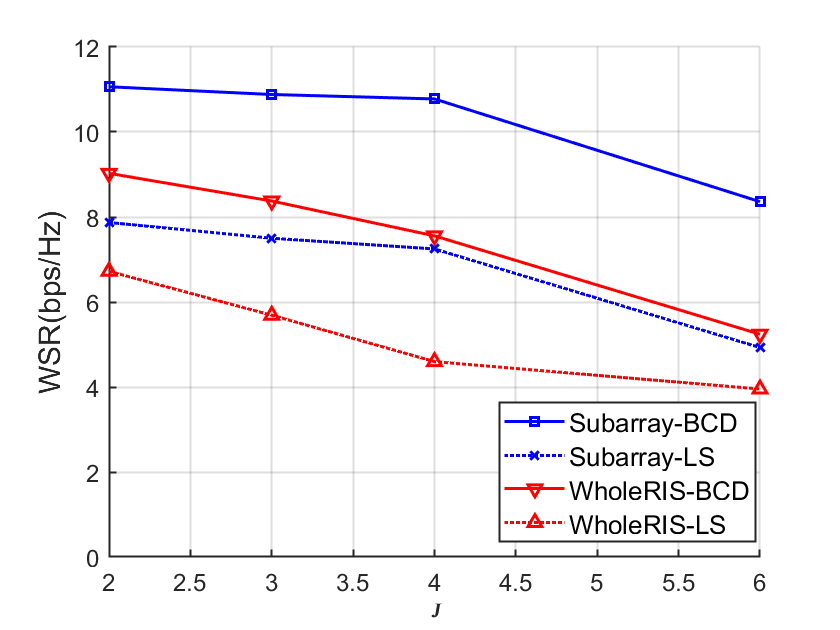}
    \caption{WSR versus the number of users {$J$} (SNR=5)}
    \label{fig:WSR-Jl}
\end{figure}

Research is also conducted on the influence of the number of users against the system performance. As shown in Fig. \ref{fig:WSR-Jl}, when {$N^{2}=144$}, the WSR performance of  the subarray RIS system is much better than the one for the system without it using the same joint beamforming scheme. %It should be addressed that each RIS subarray should has the same number of reflecting antennas, and the number of users {$J$} is much lower than {$N^2$}. 
Each RIS subarray should have the same number of reflecting antennas, and the number of users, {$J$}, should be much less than {$N^2$}.

\section{Conclusion}
In this study, we proposed a dynamic RIS subarray structure where the subarray amount is flexible according to the number of users. Besides, a WMMSE-LS scheme with limited RIS phase shift accuracy and an adaptive BCD-aided joint beamforming approach were proposed to improve throughput performance for the investigated RIS THz system. Various system models and joint beamforming methods were compared in this research. The simulation results demonstrated that the proposed dynamic RIS subarray structure combined with the adapted BCD algorithm can improve the system performance of the RIS-aided THz MIMO communication system. In our future work, machine learning based resource allocation, such as, \cite{leng2022FL,Nguyen2022FL}, for our proposed RIS systems will be considered.

% indicate that despite the fact that both the adapted BCD algorithm and the dynamic RIS subarray structure can improve the system performance of the RIS-aided THz MIMO communication system, the former has a better system performance under the same conditions while the latter has a lower computational complexity. Based on that, the proposed system with both dynamic RIS subarray structure and the adapted BCD-aided algorithm, which combines the advantages of both, will have a better system performance.

% \bibliographystyle{IEEEtran}
% \bibliography{reference}

\end{document}